\documentclass[aip, apl, superscriptaddress, reprint]{revtex4-1}

\usepackage{siunitx}
\usepackage{chemformula}
\usepackage{graphicx}
\usepackage{hyperref}
\usepackage{verbatim}

\newcommand{\mr}[1]{\mathrm{#1}}

\newcommand{\cf}[0]{cf.\ }
\newcommand{\ie}[0]{i.e.\ }

\newcommand{\fref}[1]{Fig.~\ref{fig:#1}}

\newcommand{\eref}[1]{Eq.~(\ref{eq:#1})}
\newcommand{\Cref}[1]{Chapter~\ref{chap:#1}}
\newcommand{\cref}[1]{Ch.~\ref{chap:#1}}

\newcommand{\Vanl}[0]{V_\mr{nl}}
\newcommand{\Valoc}[0]{V_\mr{loc}}

\begin{document}

\title{Focused ion beam modification of non-local magnon-based transport in yttrium iron garnet/platinum heterostructures}

\author{Richard Schlitz}
\email{richard.schlitz@tu-dresden.de}
\affiliation{Institut f{\"u}r Festk{\"o}rper- und Materialphysik, Technische Universit{\"a}t Dresden, 01062 Dresden, Germany}
\author{Toni Helm}
\affiliation{Max Planck Institute for Chemical Physics of Solids, 01187
Dresden (Germany)}
\affiliation{Dresden High Magnetic Field Laboratory (HLD-EMFL),
Helmholtz-Zentrum Dresden-Rossendorf, 01328 Dresden, Germany}
\author{Michaela Lammel}
\affiliation{Leibniz Institute for Solid State and Materials Research Dresden (IFW Dresden), Institute for Metallic Materials, 01069 Dresden, Germany}
\author{Kornelius Nielsch}
\affiliation{Leibniz Institute for Solid State and Materials Research Dresden (IFW Dresden), Institute for Metallic Materials, 01069 Dresden, Germany}
\affiliation{Technische Universit{\"a}t Dresden, Institute of Materials Science, 01062 Dresden, Germany}
\author{Artur Erbe}
\affiliation{Helmholtz-Zentrum Dresden-Rossendorf e.V., Institute of Ion Beam Physics and Materials Research, 01328 Dresden, Germany}
\author{Sebastian T. B. Goennenwein}
\affiliation{Institut f{\"u}r Festk{\"o}rper- und Materialphysik, Technische Universit{\"a}t Dresden, 01062 Dresden, Germany}
\date{\today}

\begin{abstract}
    We study the impact of \ch{Ga} ion exposure on the local and non-local magnetotransport response in heterostructures of the ferrimagnetic insulator yttrium iron garnet and platinum.
	In particular, we cut the yttrium iron garnet layer in between two electrically separated wires of platinum using a \ch{Ga} ion beam, and study the ensuing changes in the magnetoresistive response.
    We find that the non-local magnetoresistance signal vanishes when the yttrium iron garnet film between the Pt wires is fully cut, although the local spin Hall magnetoresistance signal remains finite.
    This observation corroborates the notion that pure spin currents carried by magnons are crucial for the non-local magnetotransport effects observed in magnetic insulator/metal nanostructures.
\end{abstract}

\maketitle

Pure spin transport phenomena give access to important magnetic and magnonic properties of magnetic insulators.\cite{Althammer:2013, Nakayama:2013, Chen:2013, Aqeel:2015, Cornelissen:2015, Goennenwein:2015, Ganzhorn:2016, Hoogeboom:2017}
Taking advantage of the spin Hall effect and the inverse spin Hall effect, pure spin currents can be driven into and detected across the interface between a spin Hall active metal and a magnetic insulator.\cite{Hirsch:1999, Chen:2013,Althammer:2013, Zhang:2012:1, Zhang:2012:2} 
In particular, the spin Hall magnetoresistance (SMR) allows to probe the magnetic sublattice structure of the magnetic insulator.\cite{Althammer:2013, Nakayama:2013, Chen:2013, Aqeel:2015, Ganzhorn:2016, Hoogeboom:2017}
In addition, non-local experiments in nanoscale structures with several electrically separated metal wires allow to experimentally access the magnon diffusion length in the magnetic insulator.\cite{Zhang:2012:1, Zhang:2012:2, Cornelissen:2015, Goennenwein:2015, Bender:2015}
This non-local approach thus enables studies of the properties of magnetic excitations and their diffusion in the ferromagnetic insulator, and holds the potential for faster and more energy-efficient data processing applications.\cite{Chumak:2014, Ganzhorn:2016:MBL}

\begin{figure}[h!]
    \includegraphics[width=\columnwidth]{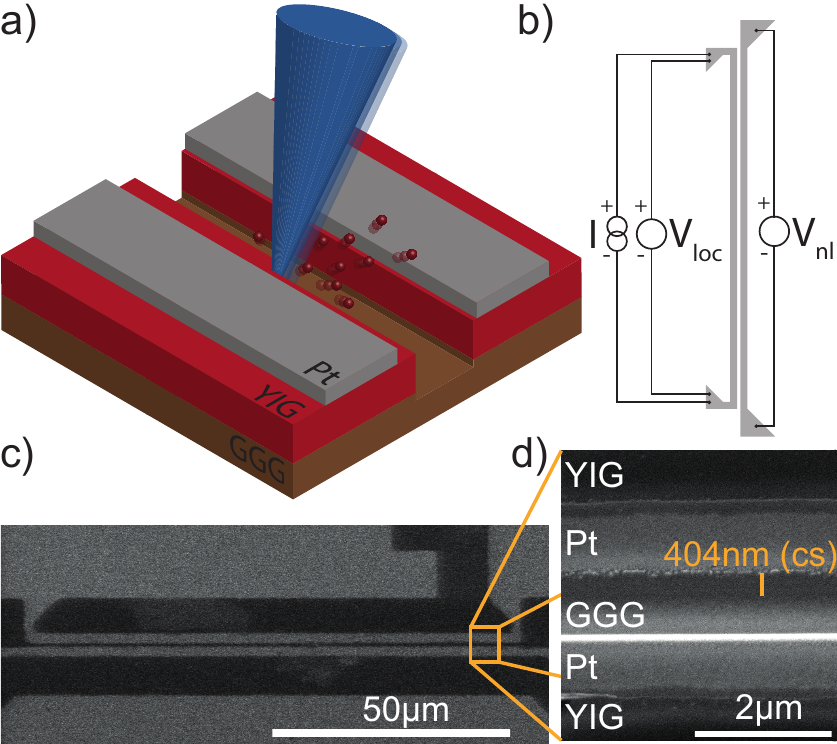}
    \caption{\label{fig:rot}
	a) Concept sketch of the experiment: The yttrium iron garnet (red) in between the two platinum wires (grey) is removed with a focused ion beam.
	b) Contacting scheme for the local and non-local magnetotransport measurement and the corresponding polarities.
	c) Scanning ion microscope image of the patterned sample. 
	The angle of incidence during the imaging was perpendicular to the sample plane. 
	The light regions are the platinum wires and bond pads, the darker regions are the YIG film.
	d) Scanning electron microscope image of the patterned sample after being fully cut down to the GGG substrate by the focused ion beam. 
	The image was acquired under an angle of \SI{52}{\degree} and shows that the cut is roughly \SI{400}{\nano\meter} deep.
    }
\end{figure}

While the non-local transport experiments in magnetic insulator/metal heterostructures so far were discussed and modeled in terms of magnon diffusion, a smoking gun proof that magnons indeed are the relevant transport quantum has not been put forward. 
Moreover, speculations about different mechanisms surfaced recently, suggesting that angular momentum can also be carried by phonons, that couple the lattice to the magnetic system.\cite{Streib:2018}

Additionally, the observation of non-local transport signatures in Pt wires on a paramagnetic substrate cast doubt on the importance of magnons.\cite{Oyanagi:2018}
Furthermore, higher order effects attributed to magnon swasing and condensation have been observed.\cite{Wimmer:2018, Thiery:2018, Cornelissen:2018, Bender:2014}
In order to ascertain that magnetic excitations indeed are essential for non-local magnetotransport in magnetic insulator/metal heterostructures, it thus appears mandatory to perform a ``scratch test", \ie, to remove the magnetic material in between the metal wires (\cf \fref{rot}a)).

In this letter we use a focused ion beam (FIB) of \ch{Ga} ions to alter and remove the yttrium iron garnet (YIG) in between two platinum wires nanopatterned onto the YIG surface. 
By performing this nano-scale scratch test, we verify that the non-local effects indeed are suppressed when the YIG is removed.
Our findings thus corroborate the notion that magnon diffusion is key for the non-local magnetotransport in magnetic insulator/metal heterostructures.

The samples were prepared starting from commercially available, \SI{180}{\nano\meter} thick YIG (\ch{Y3Fe5O12}) films deposited on (111)-oriented \ch{Gd3Ga5O12} (GGG) substrates via liquid phase epitaxy.
Prior to the platinum deposition, the samples were first cleaned using a Piranha solution (\ch{H2SO4:H2O2} in a ratio of 1:1 by volume) to remove organic constituents from the YIG surface. 
They were then annealed at \SI{200}{\celsius} for \SI{1}{\hour} in vacuum in the sputtering chamber, before a \SI{3}{\nano\meter} thick platinum film was sputter deposited.
The Pt films were subsequently patterned via optical lithography and \ch{Ar^+} ion milling. 
The typical distance between our platinum wires is $d_\mr{NL} \sim \SI{1}{\micro\meter}$ and their length and width is $l_\mr{Pt} \sim \SI{100}{\micro\meter}$ and $w_\mr{Pt} \sim \SI{2}{\micro\meter}$, respectively.

To obtain the local and non-local magnetoresistive response, we use a Keithley 2450 sourcemeter to drive a DC current of $I = \SI{200}{\uA}$ along one platinum wire, referred to as the left wire for simplicity in the following. 
The local $V_\mr{loc,raw}$ and non-local $V_\mr{nl,raw}$ voltage drop is simultaneously detected by two Keithley 2182 nanovoltmeters on the left and right wire, respectively (\cf\ \fref{rot}b)).
To increase the measurement sensitivity and to allow the separation of effects related to electric and thermal magnon generation, we employ a current reversal technique:\cite{Goennenwein:2015} 
\begin{equation}
	V_i = \frac{V_{i,\mr{raw}}(+I) - V_{i,\mr{raw}}(-I)}{2}
	\label{eq:asym}
\end{equation}
In particular, this antisymmetrization allows to investigate only the resistive response (as opposed to the thermal response observed in the symmetric signal).\cite{Avci:2015, Schreier:2013}
A magnetic field $\mu_0 H = \SI{1.1}{\tesla}$ is applied using a cylindrical Halbach array\cite{Halbach:1980}. 
By rotating this diametrically magnetized Halbach array (the angle of rotation is denoted as $\alpha$), we can measure the magnetotransport response as a function of the magnetic field orientation. 

The FIB patterning and the scanning ion and electron microscopy experiments were conducted in a FEI Helios NanoLab DualBeam system. 
The aperture for the \ch{Ga} FIB was set to \SI{0.34}{\nano\ampere} and the acceleration voltage was \SI{30}{\kilo\volt}. 
A scanning ion image of a typical structure prior to cutting and an electron microscopy image depicting the structure after FIB patterning are shown in \fref{rot}c) and d), respectively.

We will now turn to the discussion of the transport signatures.
Prior to FIB milling, the local and non-local transport was characterized.
The resulting local and non-local signal $\Valoc$ and $\Vanl$ are shown in \fref{s4comp}a).
We obtain the magnetoresistance (MR) from the raw data by using
\begin{equation}
	\mr{MR} = \frac{\Delta\Valoc(\alpha)}{V^0_\mr{loc}} = \frac{\Valoc(\alpha) - \min(\Valoc)}{\min(\Valoc)}.
\end{equation}
This MR exhibits a $\sin^2(\alpha)$ modulation and was verified to show the signature of the spin Hall magnetoresistance (SMR) from rotations in three mutually orthogonal rotation planes.
For further details on the fingerprint of the SMR, please refer to refs. \cite{Althammer:2013, Goennenwein:2015}.
The effect magnitude $\max(\mr{MR})= \SI{5e-4}{}$ (\ie the amplitude of the $\sin^2(\alpha)$ modulation) is within the typical range of values reported for YIG/Pt bilayers with \SI{3}{\nano\meter} of Pt deposited ex-situ via sputtering.\cite{Puetter:2017, Schlitz:2018}

The non-local signal also shows a $\sin^2(\alpha)$ modulation with an amplitude of \SI{1.2}{\micro\volt} and no offset. 
This is the behavior expected for the magnon mediated magnetoresistance (MMR).\cite{Goennenwein:2015, Cornelissen:2015}
Please note, that the non-local voltage shown in \fref{s4comp}a) has only been antisymmetrized (\cf \eref{asym}), no offset has been subtracted.

After FIB-cutting the YIG film in between the two platinum wires, removing a square with an area of \SI{120}{\micro\meter} and a width of \SI{500}{\nano\meter} to a depth of \SI{400}{\nano\meter}, the magnetotransport response changes significantly (\cf \fref{s4comp}b)).
While the SMR decreases by a factor of 5 to $\max(\mathrm{MR}) \approx \SI{1e-4}{}$, the MMR (\ie\ the modulation of the non-local voltage with magnetic field orientation) vanishes to within the experimental resolution ($\sim \SI{5}{\nano\volt}$) and thus is reduced by at least a factor 200.
Please note, that the slight positive offset of \SI{130}{\nano\volt} is most likely a result of a cross conductivity between the two Pt strips caused by \ch{Ga} ion implantation. 
Assuming that the (anisotropic) magnetoresistance visible in the conductive channel is \SI{1}{\percent}, the cross conduction will lead to a maximum modulation of \SI{1}{\nano\volt}, well below our resolution limit.

The absence of a non-local magnetotransport signal could be naturally explained if the right (``non-local") Pt wire would have been destroyed or made spin Hall inactive by the FIB process.
We therefore repeated the magnetotransport experiments now using the right wire as the ``local" wire, and detecting the non-local voltage on the left wire (\cf \fref{s4comp}c)). 
The resulting measurements show the same salient features, \ie\ a clear local SMR also on the right Pt strip, along with a complete absence of the MMR. 
Again, the local SMR amplitude is reduced as compared to the measurement taken before the FIB cut.
This demonstrates that both Pt wires still are spin Hall active, with a clear local SMR response. 
Since, the local SMR and the non-local MMR depend on electrically generated and detected spin transport, they both are sensitive to the square of the spin Hall angle. 
Thus, a reduction of the spin Hall angle of the platinum films due to the FIB process can not account for the vanishing non-local effect.\cite{Cornelissen:2015, Zhang:2012:1, Zhang:2012:2, Chen:2013, Althammer:2013}
These findings suggest that magnon transport indeed is mandatory for a finite MMR. 
The FIB cut removes the magnetic material in between the Pt wires and thus removes the magnetic medium mandatory for magnon transport.
\onecolumngrid

\begin{figure}[th]
    \includegraphics[width=\columnwidth]{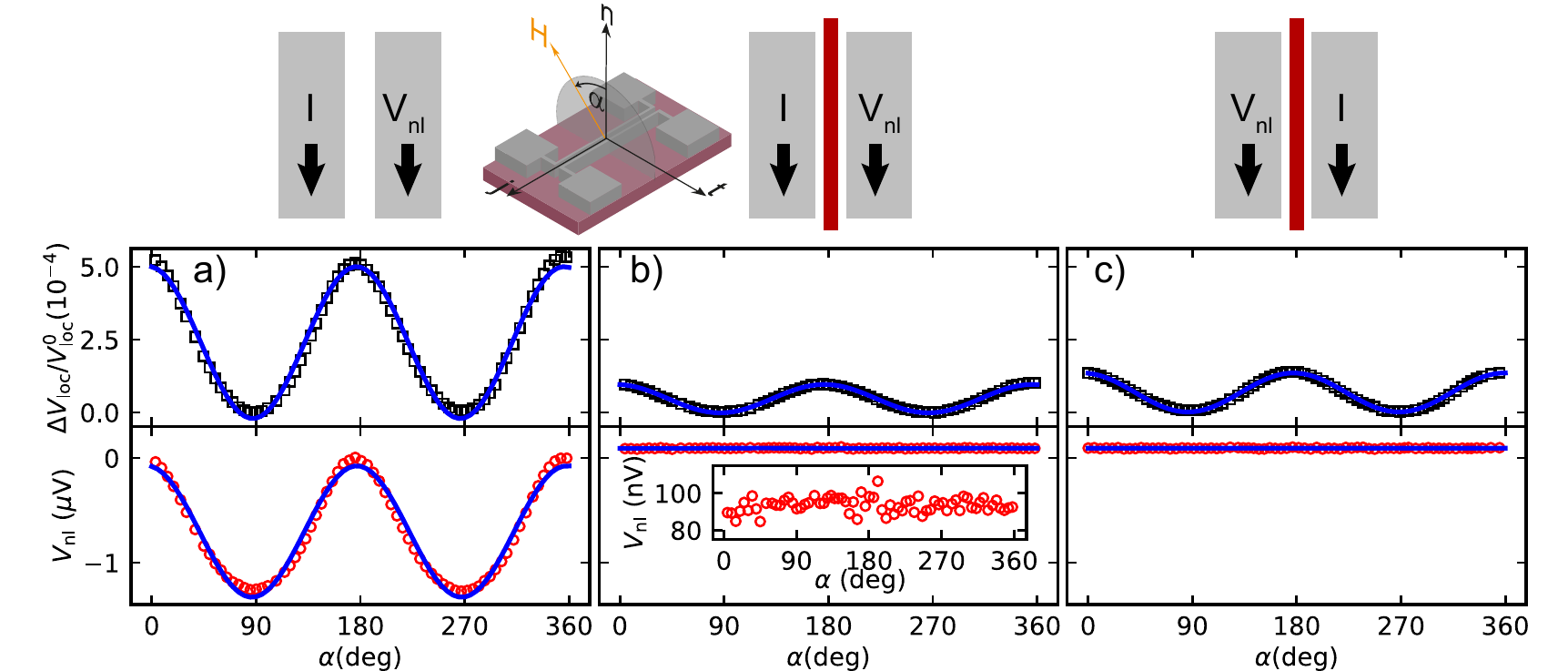}
    \caption{\label{fig:s4comp} 
	a) Local (upper panel) and non-local (lower panel) transport response prior to altering the sample with the focused ion beam.
	A clear $\sin^2(\alpha)$ signal is observed for both configurations, as expected for spin Hall magnetoresistance and its non-local analogon.
	b) After removing the YIG between the Pt wires, the local signal decreases by roughly a factor of 5. 
	The non-local signal vanishes to within our experimental resolution, while a small constant offset voltage appears.
	c) Same measurement as b) with inverted contacts (\ie current is driven on the right strip and non-local voltage is detected on the left strip). 
	Again a clear local signal is observed, while no non-local signal can be detected to within our experimental accuracy.
	The inset shows the plane of rotation of the magnetic field.
    }
\end{figure}
\twocolumngrid
Apart from magnon diffusion, another mechanism to consider for a finite non-local signal is the transport of the angular momentum by phonons.
Since phonons can propagate in particular also in the substrate and not only in the YIG film, this transport channel for the non-local voltage generation should remain open even after FIB patterning of the YIG.
Please note, that we do increase the diffusion distance when removing material in between the platinum wires: 
Upon removal of a square cross section with a width of \SI{500}{\nano\meter} and a depth of roughly $\sim\SI{500}{\nano\meter}$, the effective diffusion distance is increased approximately by a factor of 2 (\ie from 1 to \SI{2}{\micro\meter}).
Assuming a diffusion length of $0.5$ to \SI{1}{\micro\meter} and one dimensional diffusion, the signal would be reduced by a factor of $\exp(-1/0.5)/\exp(-2/0.5) \approx 7$ or $\exp(-1)/\exp(-2) \approx 3$, respectively.\cite{Cornelissen:2015, Goennenwein:2015}
This cannot explain our findings, even if we include the reduction of the local signal by a factor of 5.
Therefore, we conclude that indeed magnons must be the transport quantum relevant for the non-local magnetotransport effect.

To further elucidate the involved mechanisms, we used a second device and performed a cut over only half the length of the platinum wire.
The resulting non-local magnetoresistance measurements before and after cutting are shown in \fref{s12comp}a) and b), respectively.
In a simple picture, one would expect that the removal of half of the YIG in between the platinum wires would lead to a factor of 2 reduction of the non-local signal.
In our experiment, however, the MMR is decreased by roughly a factor of three.
Please note, that for this device, in contrast to the reduction to 0.2 of the initial value for the full cut device, here, the SMR is only decreased to 0.6 of the value prior to cutting.

We speculate, that either the YIG/Pt interface or the layers themselves are modified already by imaging of the device with the scanning ion beam, explaining the reduction of the magnetotransport response beyond the factor of 2 expected from the geometrical changes (YIG cut half-length). 
Since the penetration depth of the \ch{Ga} ions at the given acceleration voltage is in the range of $\sim \SI{10}{\nano\meter}$, this assumption is reasonable.\cite{Moll:2018}
However, more systematic \ch{Ga} ion beam irradiation experiments will be required in the future to clarify this point.
Regarding the non-local signal the experimentally observed reduction of the MMR by a factor of three can be straightforwardly rationalized as follows:
Considering, that half of the device can not contribute (no YIG film in between the Pt strips), the non-local signal in the functioning half of the device is decreased to roughly 0.7 of the inital amplitude upon irradiation.
We thus must conclude that the ion irradiation alone cannot be sufficient to explain the absence of the non-local magnetotransport signal in the first device (\cf \fref{s4comp}).

Finally, to also study the impact of the increased \ch{Ga} ion irradiation between the platinum wires, we investigated a third device, where the FIB cut was performed over the full length of the device but with a lower depth of 50-\SI{70}{\nano\meter}, \ie less than half of the YIG thickness.
The corresponding non-local magnetoresistance curves prior and after cutting are shown in \fref{s12comp}c) and d).

\onecolumngrid

\begin{figure}[th]
    \includegraphics[width=\columnwidth]{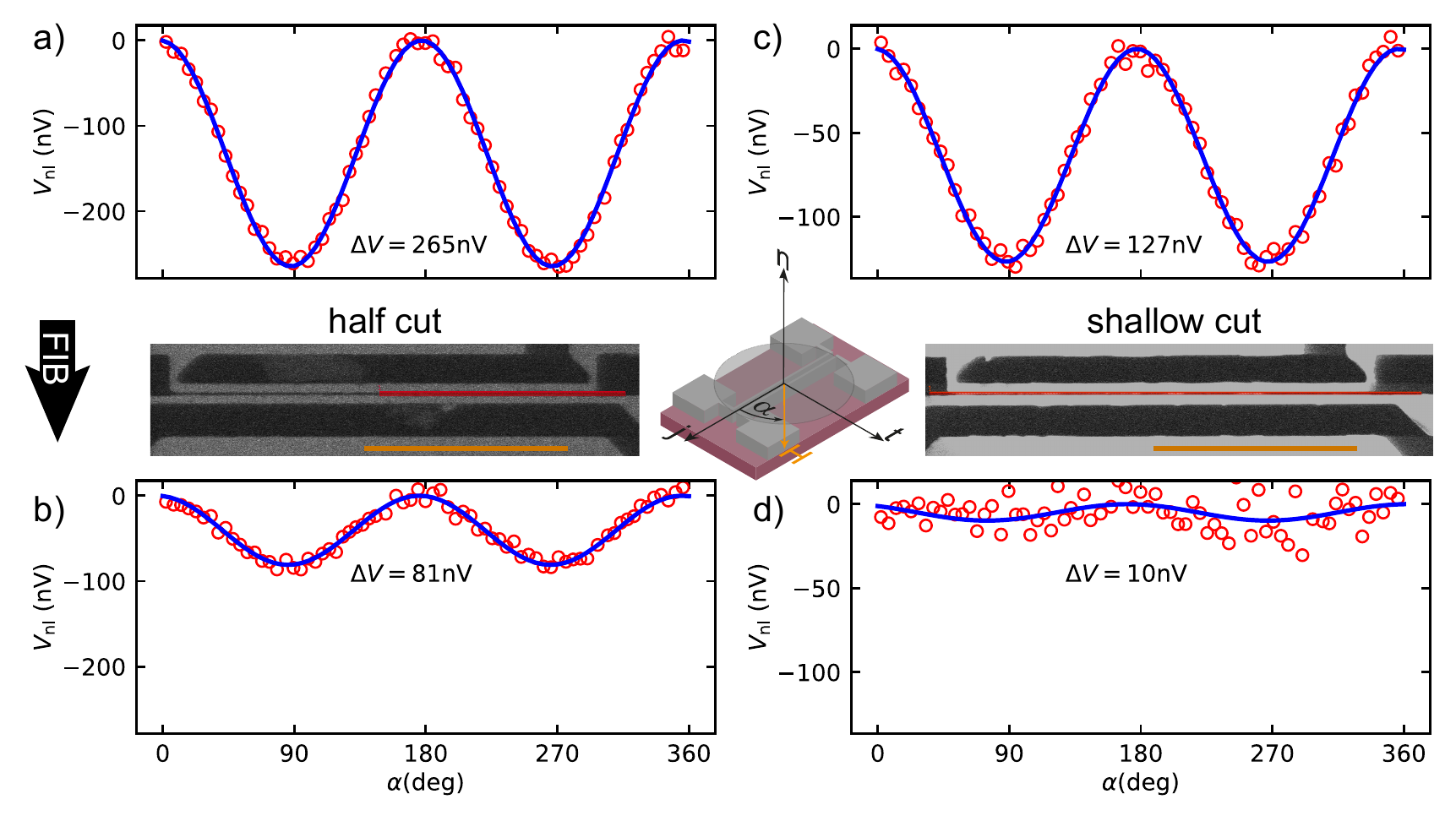}
    \caption{\label{fig:s12comp}
	Non-local magnetotransport measurement of a device before a) and after b) removal of half of the YIG in between the wires. 
	The signal amplitude goes down by a factor of 3 although only half of the length was cut away.
	Panels c) and d) show similar measurements on a device where a shallow (\SI{50}{\nano\meter} deep) cut was performed on the full length. 
	Here, the MMR is reduced by roughly a factor of 10.
	The regions marked in red in the scanning ion microscopy images in between the plots are the regions that were cut away using the focused ion beam.
	The orange scale bars have a length of \SI{50}{\micro\meter}.
	An offset $\le \SI{1}{\micro\volt}$ has been removed from all curves to allow easier comparison of the data.
	The central inset illustrates the magnetotransport experiment.
    } 
\end{figure}
\twocolumngrid

We will now try to extrapolate the non-local magnetrotransport signal based on our previous observations:
We expect that the signal is decreased to about 0.7 of its initial value just because of the imaging of the device with the \ch{Ga} ions.
Additionally, we have removed roughly half of the YIG layer (thickness) in the channel with the FIB. 
Thus, the signal should be reduced to roughly $\SI{127}{\nano\volt} \cdot 0.7 \cdot 0.5 = \SI{44}{\nano\volt}$, which is approximately four times larger than the signal observed in the experiment (\cf \fref{s12comp}d)).
We therefore conclude that either \ch{Ga} ion implantation is very efficient in altering the magnonic properties of the YIG film or the constriction of the transport channel leads to additional boundary conditions for the magnon transport.\cite{Chumak:2017}
This can be potentially useful in the creation of magnonic crystals necessary for single mode magnon transport, where periodic alterations of the YIG properties are crucial.\cite{Chumak:2017}
Please note, that also for the shallow cut, the SMR is reduced only to 0.6 of its initial value, compared to 0.2 for the deep cut. 

In conclusion, we have found that FIB patterning of the YIG film in between the two platinum wires very sensitively affects the local and non-local magnetotransport response of YIG/Pt heterostructures.
Our results corroborate the notion that magnon diffusive transport is key for the observed non-local magnetotransport signatures in these nanostructures.
Additionally, by studying the impact of FIB cuts to different depths and with different lengths, we conclude that YIG/Pt devices are very sensitive to \ch{Ga} ion irradiation. 
This opens new avenues for (periodic) modulation of the magnonic/magnetic properties of such devices.

We thank T. Sch{\"o}nherr and S. Piontek for technical support. We acknowledge financial support by the Deutsche Forschungsgemeinschaft via SPP 1538 (project no.\ GO 944/4).

\bibliography{bibliography_rs_v01}
\end{document}